\documentclass[5p,times]{elsarticle}

\usepackage{xcolor}
\usepackage{amssymb}
\usepackage{amsmath}
\usepackage{soul}
\usepackage{nomencl}
\makenomenclature
\begin{document}

\begin{frontmatter}



\title{\textsc{TALICS$^3$}: 
Tape Library Cloud Storage System Simulator}

\author[1,2,3]{Suayb S. Arslan}
%\cormark[1]


\ead{sarslan@mit.edu, arslans@mef.edu.tr, suayb.arslan@quantum.com}

\ead[url]{www.mit.edu/~sarslan}

%\credit{Conceptualization of this study, Methodology, Software}

\affiliation[1]{organization={Massachusetts Institute of Technology},
    addressline={77 Mass. Ave.}, 
    city={Cambridge},
    % citysep={}, % Uncomment if no comma needed between city and postcode
    postcode={02139, MA}, 
    country={USA}}

%\author[2]{Han Theh Thanh}[style=chinese]

\author[2]{James Peng}
\ead{james.peng@quantum.com}
%\ead[URL]{www.sayahna.org}

%\credit{Data curation, Writing - Original draft preparation}

\affiliation[2]{organization={Quantum Corporation},
    city={Irvine},
    % citysep={}, % Uncomment if no comma needed between city and postcode
    postcode={}, 
    state={CA},
    country={USA}}

\author%
[2]
{Turguy Goker}
%\cormark[2]
%\fnmark[1,3]
\ead{turguy.goker@quantum.com}
%\ead[URL]{www.stmdocs.in}

\affiliation[3]{organization={MEF University},
   addressline={Huzur Mah. Maslak}, 
    city={Istanbul},
    % citysep={}, % Uncomment if no comma needed between city and postcode
    postcode={34396}, 
    %state={},
    country={Turkey}}

\cortext[cor1]{Corresponding author.}
%\cortext[cor2]{Principal corresponding author}

\fntext[fn1]{Both the first and the second authors eqaully contributed to the development of the software.}


\begin{abstract}
High performance computing data is surging fast into the exabyte-scale world, where tape libraries are the main platform for long-term durable data storage besides high-cost DNA. Tape libraries are extremely hard to model, but accurate modeling is critical for system administrators to obtain valid performance estimates for their designs. This research introduces a discrete--event tape simulation platform that realistically models tape library behavior in a networked cloud environment, by incorporating real-world phenomena and effects. The platform addresses several challenges, including precise estimation of data access latency, rates of robot exchange, data collocation, deduplication/compression ratio, and attainment of durability goals through replication or erasure coding. Using the {proposed} simulator, {one can} compare the single enterprise configuration with multiple commodity library configurations, making it a useful tool for system administrators and reliability engineers. This makes the simulator a valuable tool for system administrators and reliability engineers, enabling them to acquire practical and dependable performance estimates for their enduring, cost-efficient cold data storage architecture designs. %They can use the simulator to obtain practical and reliable performance estimates for their long-term, durable, and cost-effective cold data storage architecture designs.
\end{abstract}


\begin{highlights}
\item A realistic (discrete--event) tape library-based cloud system simulator is proposed.
\item In the simulator, one can employ multiple libraries to provide distributed archival backend. 
\item An analysis on read-latency has been conducted to motivate for the simulator.
\item Comparative analyses evaluate distributed systems v.s. centralized single-library systems.
\item The presented simulation platform serves as a valuable design tool for reliability engineers.
%A realistic (discrete--event) tape library-based cloud system simulator is proposed with various parameters to adjust based on the use cases, architectural decisions and environmental conditions. 
%\item The scope of the simulator has been expanded to include distributed environments that employ multiple libraries to provide cloud-based data storage with an archival backend. Comparative analyses are conducted to evaluate these distributed systems in relation to centralized single-library systems.
%\item An investigation on read-latency has been conducted to demonstrate the limitations of using closed-form performance equations, highlighting the need for the implementation of the simulator as a more practical alternative. 
\end{highlights}
\end{frontmatter}

\nomenclature{$C_t$}{Tape Capacity (MB)}
\nomenclature{$\Phi_f$}{Library fill ratio}
\nomenclature{AOTR}{Annual Object Touch Rate}
\nomenclature{\textit{NoC}}{The number of cartridges}
\nomenclature{\textit{NoT}}{The number of objects touched}
\nomenclature{xph}{Exchanges per hour}
\nomenclature{$\mu_o$}{Average Object size}
\nomenclature{$m_i$}{Mean size of the requested data by user $i$}
\nomenclature{$\lambda$}{Parameter of Poisson arrivals}
\nomenclature{$k$}{The number of equal-size raw data fragments}
\nomenclature{$t$}{Data fragment requests that never arrive}
\nomenclature{$n$}{Codeword size}
\nomenclature{$s$}{Number of fragment requests}
\nomenclature{$\mathcal{N}_n$}{Ordered set of integers up to $n$}. 
\nomenclature{$\mathcal{S}_k$}{Any $k$-size subset of $\mathcal{N}_n$}. 
\nomenclature{$p_d$}{Drive failure probability}
\nomenclature{$N$}{Number of library nodes}
\nomenclature{$L$}{Number of users}
\nomenclature{$l$}{Number of failed cartridges or libraries}
\nomenclature{$c$}{Number of robots ($r$)/drives ($d$)}  
\nomenclature{$L_q$}{mean number of requests awaiting service}
\nomenclature{$P_0$}{Probability that there are no requests waiting in the queue}
\nomenclature{$W_q/G_q$}{Wait time in the queue}
\nomenclature{$C_a^2$}{Coefficients of variation of the inter-arrival time}
\nomenclature{$C_s^2$}{Coefficients of variation of the service time}
\nomenclature{$s_R$}{Mean robot service time}
\nomenclature{$s_D$}{Mean drive service time}
\nomenclature{KPI}{Key Performance Indicator}
\nomenclature{RAIL}{Redundant Array of Independent Libraries}
\nomenclature{LTFS}{Linear Tape File System}
\nomenclature{ML}{Multi Library}

{
\printnomenclature}

%\maketitle

\section{Introduction}

With an unprecedented historical demand for data, we encounter a rapid surge in our need for digital content storage. This includes a remarkable rise in the volume of sporadically accessed data, characterized by significantly extended retention periods, contributing to our escalating storage demands. {Additionally, the emergence of generative AI has further fueled this data explosion, adding yet another source of data proliferation.} International Data Corporation (IDC) had accurately estimated that 161 exabytes of digital information were produced in 2006 and recently projected the production of nearly 175 zettabyte new data in 2025 \citep{IDC2025}. As it is clear that the same trends will dominate the aggregate data growth in the world, there is enough evidence that we will soon be dealing with few hundreds of zettabytes \citep{gantz2011extracting}. Given this observation, it becomes crucial to consider the cost implications linked to durable, long-term storage of data--intensive applications. In this context, tape technology undoubtedly stands out as a cost-effective nearline alternative, offering one of the most economical cost of ownership. For instance, tape is estimated to be three times cheaper than HDDs in popular data centers \citep{Moore} and it is way cheaper compared to today's high density DNA storage technology due to the high cost of read/write processes of DNA using expensive \textit{synthesis} and \textit{sequencing} operations \citep{mardis2011decade, Appuswamy2019, Bornholt2016}. 

On the other hand, data-intensive scientific domains such as High Performance Computing (HPC) serve as significant sources for swiftly generating substantial {amount of} data volumes. Scientific domains vary considerably in data access patterns, storage capacity, and specific requirements, contingent upon the particular application in use\footnote{These applications include earth observations, radio astronomy, nuclear physics and medicine-related research.}. For instance, active archives require the data to be online all the time whereas in static archives, written bulk data is only rarely accessed and modified \citep{gupta2012enabling}. The decision-making process for selecting the appropriate media, robot technologies, library configurations, drives, communication links, and software can be heavily influenced by the specific requirements of the application. Hence, there can be significant variations in the chosen components and their configurations depending on the nature and demands of the performed task. Recently, several cloud service providers have started offering Cold-Storage-as-a-Service (CSaaS) with varying access and storage capacity options to address the requirements of long-term scientific data retention \citep{memishi}. 

Tape systems are commonly available as libraries comprising a master server, tape drives, robots, and a collection of cartridges. These cartridges, made of magnetic tapes with a capacity of $C_t$ MBs each, form the primary medium for storing extensive data in bulk within the library, organized in specific 2D/3D geometries. The drives within this system undertake the writing and reading of tape cartridges following a designated format. The robots are responsible for moving cartridges between drives and library shelves or the so called \textit{racks}. Robots vary in quality and display diverse {operational} speeds, typically quantified in exchanges per hour (xph). For instance, enterprise-grade robots, while expensive, boast superior actuation speeds. Within tape libraries, the number of robots tends to be limited due to their high manufacturing and maintenance costs.  Finally, the server functions as the system's gateway, offering an interface to the cloud (commonly utilizing a REST command set) for external access. It adapts the data format to align with tape storage requirements and oversees internal operations such as object allocation, load balancing, compression, collocation, and data protection \citep{cancio2015experiences}. 

Tape is a highly sequential media i.e., its performance is usually unacceptable for random reads and writes \citep{pease}. Although random reads/writes may be served, posititoning the heads (particularly for small size data objects), loading the cartridge etc. may leave the system in a livelock state intermittently. {Hence, the central drawback of tape, high time-to-first-byte latency \cite{7877100}, need to be modeled accurately for future potential use cases.} Some examples of tape systems include IBM TS3500 Series, Oracle StorakeTek SL8500, Spectralogic TFinity, Quantum Scalar i6000. Each of these tape library products possesses distinct characteristics and module configurations, which can manifest varying degrees of performance based on the parameter selections and data request patterns of the user. Nevertheless, while the operational characteristics and constituent components of these systems may differ, they can be represented through a shared numerical framework.

\subsection{Motivation}

Today, tape systems come with very compelling features such as low acquisition cost, backwards compatibility, scalability, high performance, longevity, high capacity and portability. The Linear Tape File System (LTFS) \citep{pease} open tape format and backward compatibility with earlier Linear Tape Open (LTO) versions, ensure that you will be able to read and restore data in the future without the need for proprietary applications.  Tape can also provide an eventual defense mechanism against malware, hacking attacks and data corruption that may affect copies of data objects stored on disk in the cloud \citep{szor2005art}. As companies begin to migrate data to cloud, tape is seriously being considered for the  archival backend for the future of web and decentralized applications such as Inter Planetary File System (IPFS), a content addressable peer-to-peer distributed file system \citep{IPFS} or digital ledger-powered surveillance applications \citep{arslan2022compress}. Yet another example is the earth observations data (D-SDA) which is operated by the Earth Observation Center (EOC) of the German Aerospace Center. The data is stored in a vast, geo-replicated cold storage data archive facility, which heavily relies on a tape library system \citep{kiemle2016}.

Despite their practicality, tape library systems encompass a range of electronics and actuation mechanisms, leading to intricate error processes that could present modeling challenges \citep{arslan2014mds}. It is essential to create precise models to estimate the overall cost for system designers addressing current cloud application requirements. Hence, this study aims to devise a concise yet effective model, aimed at quantifying critical performance metrics pivotal in hierarchical storage system design. Specifically, this model emphasizes tape systems' role as a primary archival backend. We can list some of the basic motivations below for creating such a tape simulation model. 

\begin{figure*}[!t]
\centering
\includegraphics[width=13cm,keepaspectratio]{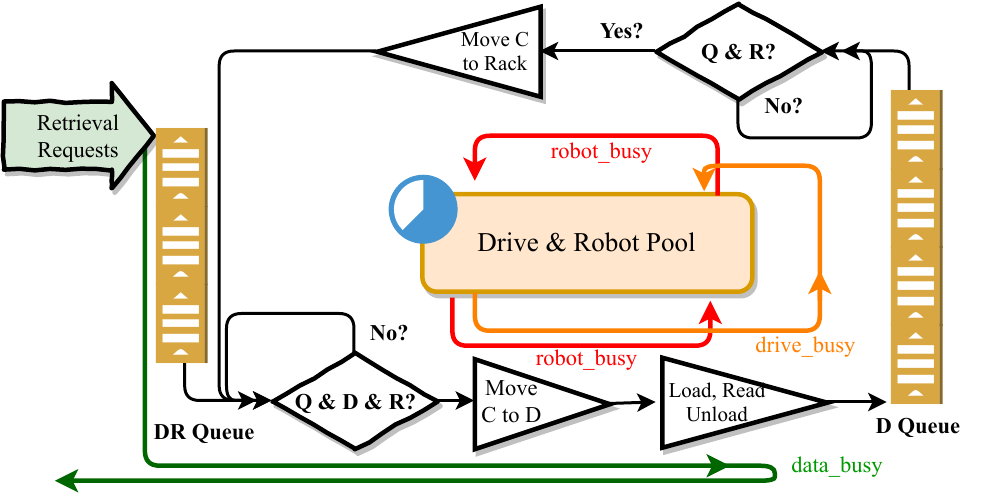}
\caption{Summary of the Proposed Simulation Architecture. \textbf{Q}: Queue, \textbf{D}: Drive, \textbf{R}:Robot, \textbf{C}: Cartridge.}
\label{fig_sim}
\end{figure*}

\begin{itemize}
    \item Due to numerous correlated dynamics of its internal design, it is very hard to model the real-world tape systems using queuing theory and statistics \citep{meisling1958discrete}. There are a few preliminary works in the literature for key performance indicators (e.g. average access latency) such as \citep{9614293}. However, their approach is not necessarily realistic and practical. In addition, not only they do not accurately capture advanced scheduling policies and asymmetric
    and bursty workloads, but also they are not easily extensible to arbitrary queuing models and inclusive of {developping new} library features. %  It will be argued that it is analytically intractable when the model gets complex w/ additional components.
    \item Utilizing a simulation platform would facilitate the implementation of modifications for added functionalities (e.g., collocation, simulating multiple libraries to mimic Redundant Array of Independent Libraries (RAIL) \citep{Ford1996}) in cloud-based tape systems. This approach would streamline the decision-making process significantly.
    %It is easier with a simulation platform to apply modifications and add extra features (ex: collocation,  multiple libraries to simulate Redundant Array of Independent Libraries (RAIL) \citep{Ford1996}) for next generation cloud-based tape systems. This would also expedite the overall decision process. 
    \item {The proposed simulator} would help us parameterize the overall system dynamics to explore all possible scenarios under different conditions. We shall be able to measure various system performance metrics which may otherwise be impossible in practice without an accurate mathematical model (e.g. robot exchange rate as a function of time). 
    \item {It would also create} enhanced understanding regarding the intricate mechanisms of real-world complex library systems. This comprehension encompasses aspects such as the {ratio} of robots {to} drives, as well as the impact of queuing models on the ultimate performance of data access latency.
    %We will obtain better insight about the inner workings of the real world complex library systems including the number of robots and drives, effect of queuing model etc. on the final data access latency performance. 
    \item A simulation platform will empower {practitioners} to simulate and analyze numerous hypothetical scenarios, allowing for comprehensive operational understanding and the formulation of precautionary measures for optimal system configuration and functionality. Additionally, this approach facilitates the identification of genuine underlying issues and enables prompt remedial actions.
    \item {Finally, the complexity of the configurations and the substantial investment required for integration render real tape systems challenging to modify and experiment with, limiting the generation of required datasets for subsequent machine learning algorithm development, particularly those (e.g., deep learning) demanding significant data volumes for achieving satisfactory accuracies. A simulator presents a potential solution for advancing the development of future smart library systems by fueling learning systems with precise and realistic data.}
    %\item {Last but not least, due to the challenging setups and costly integration, real tape systems are hard to modify and experiment with to generate necessary datasets for subsequent machine learning algorithm development, which might be data-hungy for acceptable accuracies. A simulator could be a remedy for future smart library system development fueling the hardware intelligence with accurate and realistic data.}
    %It will provide us the ability  to simulate and analyze several unrealistic and impractical scenarios to gain full operational insight and help prepare a list of precautions for the right system configuration and hence operation. Moreover, it would make it easier to identify real root-cause of problems and help take immediate action. 
\end{itemize}

\subsection{Previous work and Methodology}

In the past literature, {despite the important role the tape libraries play for voluminous long-term archives,} a few modeling attempts have been made such as \citep{9614293, Zeng1} that are either hard to configure, complex to compute, or they do not take all mechanical effects and/or advanced features of library systems into account \citep{7774566}. Consequently, they remain distant from resolving the interrelated dynamics among the library, robot, and cartridge, given their shared elements such as hardware, which foster close interactions among these system components. Creating a universal tape library model is {quite} challenging {not only} due to the difficulty in anticipating inaccuracies in the model, {but also the inability to validate its fit from the available data}. Although Markov models may be used to represent isolated and stationary failure patterns, extending them to actual usage scenarios may not be directly feasible \citep{arslan2020data}. {In \cite{masker2016simulation}, cartridge eviction strategies are analyzed through a simulation platform. It mainly focuses on deferred dismounting as the key parameter and largely ignores other important parameters such as collocation and parallel libraries.} Another alternative is proposed in \citep{tgau} that uses graph theory. However, the proposed scheme in that study is hard to configure and too complex to {run and} obtain {real-world applicable} results. %There are also 2D/3D topology models from which our model will take advantage while modeling the physical constraints of the library geometry. 
{Additionally, our model will leverage 2D and 3D topology models to incorporate the physical constraints inherent in the library geometry during the modeling process.} Although topologically correct, such models fail to address queuing requests (and capture queue delays) which is the main source of latency in tape library systems, particularly under heavy load \citep{Lavengerg}. For an introduction to reliability theory fundamentals and their application in cold data storage, readers {are kindly referred} to  \citep{arslan2023durability}. 

Discrete--Event Simulation (DES)  is the most effective way to model and simulate a wide variety of problems \citep{fishman1978}. In each discrete time instance (step) of DES, only and only one event can happen. Since such steps can arbitrarily be small, we can  simulate concurrent events almost simultaneously. This is based on the assumption that as the time steps shrink further (a configurable parameter of the simulation), the probability of seeing more than one event in a given simulation step vanishes and hence DES will be able to simulate concurrency realistically. This is particularly significant when modeling a cold archival storage system compared to hot storage models. {Previously, \cite{luttgau2016modeling} has employed DES using a graph-based topology where the drive and robot allocation and scheduling is primarily based on SL8500 library system.} In our model, we sought to incorporate all significant factors to produce the most realistic simulation output feasible while minimizing complexities by making appropriate approximations that do not considerably impact the final outcome.
Our model, {unlike previous lierature,} adapts 2D topology and takes the natural ``double queuing phenomenon" into account. For simplicity, we do not consider data communication that happens between the tape  system and the host and all associated software architectures in between.

\section{Double-Queue DES Model}

%As mentioned before, modeling a tape system is a complex task because of system components and dynamics that may work against each other. 
{As previously stated, simulating a tape system presents a challenge due to the intricate interplay of its components and dynamics, which may be at odds with one another.}
For instance, while the collocation process can help reduce the robot exchange rate, it can also lead to a more complex controller design and increases the data retrieval latency. We use DES and simplify our model as shown in Fig. \ref{fig_sim}. In this double-queue model, Data Request (DR) queue is used to place the data (file/object/fragment) requests for subsequent library service. Any data request in the DR queue is immediately processed once both a drive and a replacing robot are available in the library system configured as a pool. This is checked in every simulation step (through polling the pool).

\begin{figure}[!t]
\centering
\includegraphics[width=8.9cm,keepaspectratio]
{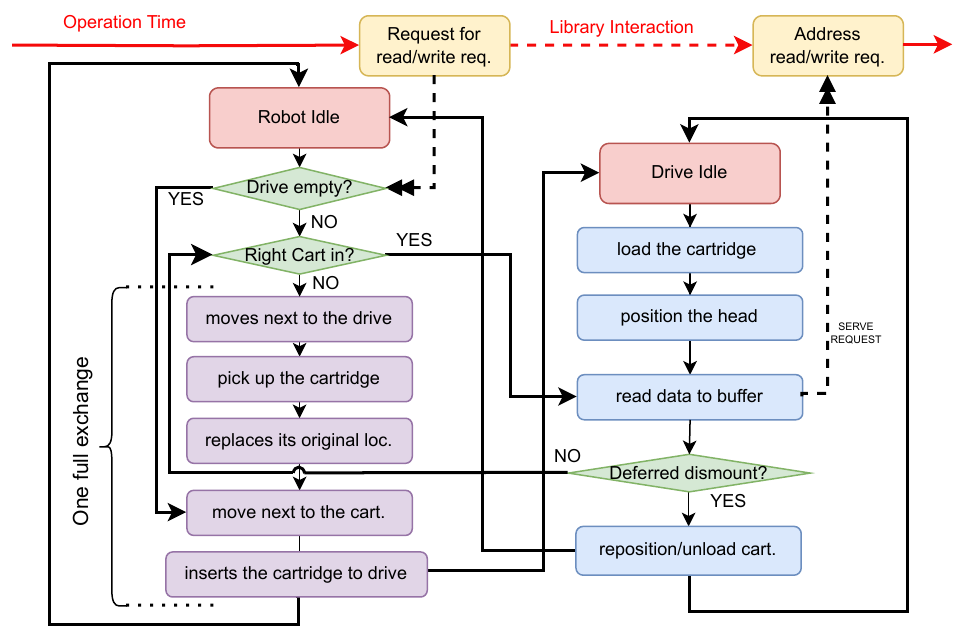}
\caption{{Schematic organigrame representation of the Operational
Cycle of Robotics and Drive Systems.} Deferred dismount is used to increase {resource} efficiency by caching the cartridge data. {As can be shown above, one full exchange consist of 5 steps: (1) Moving the robot next to an available drive, (2) Picking up the cartridge, (3) Replacing the cartridge, (4) Moving the robot next to the cartridge and (5) Inserting the corresponding cartridge into the drive.}}
\label{fig_sim_schematic}
\end{figure}

{\subsection{The operation cycle of a robot}}

The robots support cartridge movements in an operation cycle and this is repeated until there are no robots available to serve. There are three states of a library robot. Initially, the robot moves next to the drive to pick up the cartridge and replaces its original location in the library. Next, it travels for the cartridge to be inserted into the available drive. Once the new cartridge is inserted into the drive,  we consider one \textit{full exchange} has occurred. The robots that complete a full exchange get into an idle state and are recycled/returned to the available pool of drives and robots (PDRs). The drive on the other hand first loads the cartridge, positions the heads for the right location and finally reads the data to its internal buffer. The buffer is emptied once full to address the host's/client's data read request (a similar process can be described for write requests) with the help of the managing server. Once the reading process is over, the drive re-positions, unloads the cartridge, and places itself in the Drive Queue (D Queue) to be serviced by an available robot. It also frees itself from the busy state and rejoins the available PDRs once it is serviced by a robot. D Queue is serviced by available robots which move the cartridges to their respective library locations. Once done, the robots turn idle again and rejoins the available PDRs, and awaits the next job. In case there is no drive or robot available at the time of the data request, the system stalls and polls for availability in the following discrete simulation step. {Summary of these operations are outlined in {an organigrame representation} given in Fig. \ref{fig_sim_schematic} where cartridge dismount/unloading can be delayed for caching data and hence increasing efficiency}. {Additionally, deferred dismount is used to increase efficiency further by caching the cartridge
data The figure also shows what constitutes a full exchange.} {We finally note that} all queues in our simulation model are First-In-First-Out (FIFO) \cite{luttgau2016modeling} and positioning/re-positioning actions are assumed to have the same probability density functions.

{\subsection{The operation cycle of a drive}} 

Similar to robots, drives can also be in three different states. In a \textit{busy} state, drives are assigned for a service and logically reserved until they are serviced by a robot. Once the drive completes the assigned task (a reading or a subsequent writing operation), they get into an \textit{idle} state where they are queued in the D Queue to be serviced by a robot. Once the drive becomes available again it changes state to \textit{free} and is returned back to PDRs.

Data write or read requests are populated in the DR queue according to a Poisson distribution with rate $\lambda$. This rate can either be directly set or can be tied to the ratio of the data volume stored in a library relative to full raw data capacity i.e., library fill ratio ($\Phi_f$) and Annual Object Touch Rates (AOTR) which might be readily available from the past usage patterns of the library\footnote{In fact, past data can be used in association with a learning module to predict the future data access patterns.}. In our model, we can define three important measurable metrics that heavily have an impact on the performance of a library system:
\begin{itemize}
    \item[$\bullet$] \textit{data\_busy}: The time between the data gets placed on DR-Queue and the time it is served by the system after the drive completes reading.
    \item[$\bullet$] \textit{robot\_busy}: The total time robot spends moving cartridges. This may involve multiple but consecutive short movements as will be detailed later.
    \item[$\bullet$] \textit{drive\_busy}: The time the drive is released from PDRs, positions, loads, reads, unloads and re-positions the heads. In fact it also includes cartridge to cartridge and cartridge to drive robot movements as well.  Once a drive is assigned for a service, it is logically reserved, stays idle and waits for the assigned job. No drive reserved for a job can be assigned {for} another job. 
\end{itemize}

\begin{figure}[!t]
\centering
\includegraphics[width=8.9cm,keepaspectratio]
{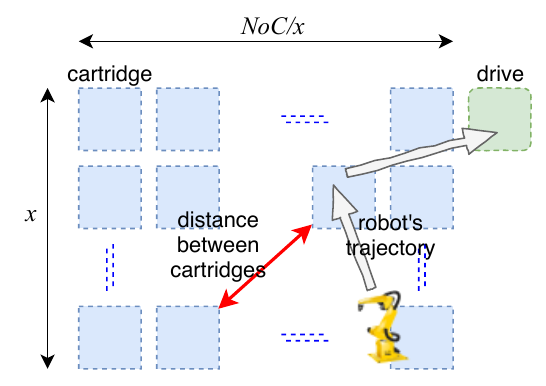}
\caption{A simple 2D {topological} model of the tape library system is presented. {The trajectory that the robot follows to transport cartridges for the drives is determined by the Euclidean distance between cartridges. The positioning of the drive is strategically placed on the upper right region of the 2D configuration, a decision guided by several optimization principles. }}
\label{fig_sim2_2dmodel}
\end{figure}

\subsection{Parameters of the System}

There are several important parameters of the system such as library geometry, robot exchange, drive dynamics and object size which shall have direct impact on the performance of the library system. {Let us explore these prominent parameters of the system in detail.}

\subsubsection{Library Geometry}
A tape library is a three-dimensional system in which drives and cartridges are placed according to a geometry. Geometry directly affects the time it takes a robot to move a cartridge from location $X$ to location $Y$ inside the library. For simplicity, we consider a two dimensional rack topology shown in rectangle shape in Fig. \ref{fig_sim2_2dmodel} which can be extended to 3D {through Euclidean geometry}. The number of cartridges (\textit{NoC}), the vertical dimension ($\textit{x}$) as well as the locations of drives are the parameters of the 2D library geometry model. In general, the number and orientation of cartridges, drives and robot movement trajectories directly affect the performance of a library. For instance, a 2D library of size $x \times NoC/x$ where a single drive is located top right and robots moving to their destination using the smallest Euclidean distance is given in Fig. \ref{fig_sim2_2dmodel}. 

Considering that the probability of being at {any} point {in a given library topology} is equally likely, the library geometry defines the distribution of travelling distance for a robot. Hence, considering a unit distance for cartridges and drives, calculation of pairwise distances can be performed. The mean value is configured/managed by the user to make the geometry resemble to the real library system topology. Finally, we emphasize that the same approach can be applied to 3D Euclidean geometry (a generic rectangular prism - cuboid) and similar mean statistics can be obtained. 

\begin{figure}[!t]
\centering
\includegraphics[width=9cm,height=7.3cm, keepaspectratio]{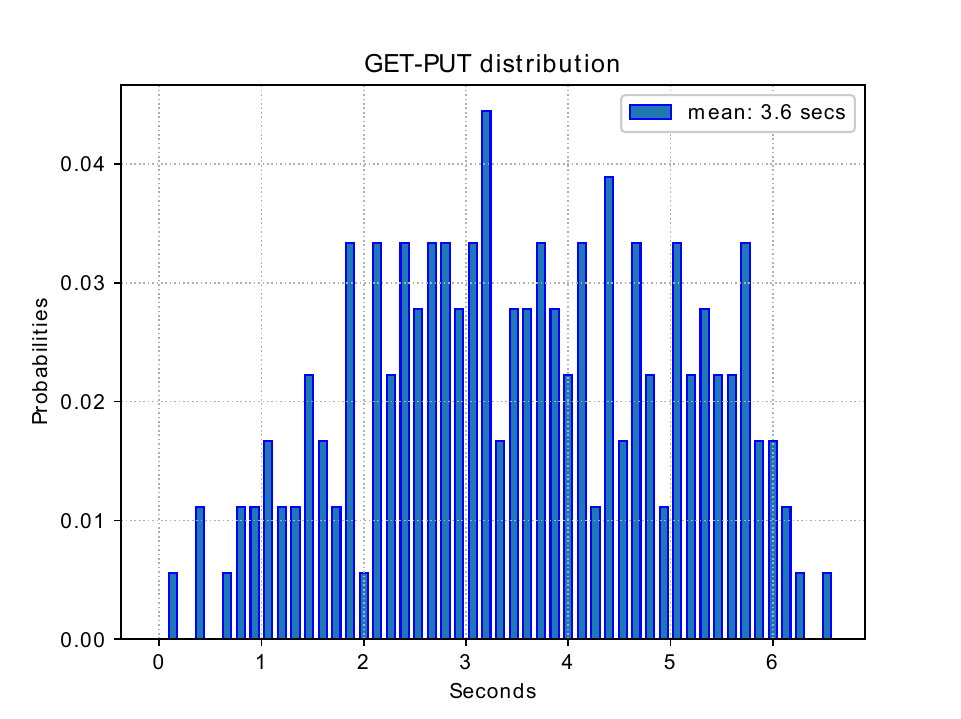}
\caption{A {time} distribution of GET-PUT operations of the robot is {presented}. These numbers do not necessarily represent realistic system performance.{All first order statistics for discrete random variables can be computed for the estimation of the GET-PUT probability distribution. In this example, the legend shows the mean value of the distribution.} }
\label{fig_sim2_dist}
\end{figure}

\textit{Robot exchange count} is defined as the number of times the robot completes a full exchange motion as defined {earlier}. %Robot exchange rate is the ratio of robot exchange count to total number of discrete simulation steps for a given period of time (e.g. an hour). In fact, maximum robot exchange rate could be considered as a wear parameter associated with each robot and  expressed in terms of exchanges per hour (xph). This is so because exceeding this maximum might lead to unreliable robot behavior. From this metric, the minimum exchange time can be calculated to be used in our simulation platform. 
{The robot exchange rate denotes the proportion of robot exchange occurrences to the total discrete simulation steps within a specified timeframe (e.g. hour). Indeed, the maximum robot exchange rate can be construed as a durability metric linked to each robot, quantified in exchanges per hour (xph) and set by the manufacturers. This characterization is essential, as surpassing this maximum threshold may result in unpredictable robot performance. Based on this metric, the minimum exchange time can be derived for utilization within our simulation platform.} In large enterprise library systems for instance, there may be more than one robot. In order to balance the wear, we pseudorandomly choose one of the robots if more than one is housed in the corresponding PDRs. One of the performance metrics that is directly tied to exchange count is the \textit{number of objects touched (NoT)} in the tape cartridges. Everytime a robot moves from cartridge to an empty drive, we increment NoT count. %In fact, this would be a lower bound on the number of objects touched if collocation is utilized.  

\subsubsection{Object Size} It is reported in various past studies  that the data object size (stored or communicated as objects with an average size of $\mu_o$) tend to possess heavy-tailed distribution such as Pareto \citep{Satyanarayanan}. On the other hand, there are also studies demonstrating that other heavy-tail distributions might be possible for local file systems \citep{downey}, hard disk data \citep{arslan2020distribution} and world wide web \citep{gong}. In our study, we assume a Weibull distribution with configurable shape and scale parameters such that we can model fixed and exponential object sizes by adjusting these parameters.

\subsubsection{Loading and Positioning} Media load time is defined as the amount of time between the cartridge insertion into a drive and the drive firmware become ready to listen for host system commands. We assume the media load time to exhibit only little variation as the drive system is an enclosed system and therefore least affected by external changes. After the cartridge is loaded into the drive, the head is being positioned on the location of the data laid on the magnetic tape. Depending on the exact data location on tape layout, the loading time may dramatically vary. We model the location of the data object to be uniformly distributed across the tape surface.

\subsubsection{Robot Exchange}

In our model, an exchange consists of four motions that can be sequenced as GET-PUT-GET-PUT. To visualize that, suppose we would like to switch to a different cartridge in a drive. Initially, we assume that robot is situated at an arbitrary point in the library and reserved for an exchange. The \textit{first motion} (r2d) consists of moving near to the drive to pick up (GET) the unloaded cartridge. In the \textit{second motion} (d2c), robot carries the cartridge to its respective location (PUT) inside the library to empty the drive for an incoming (target) one. The \textit{third motion} (c2c) of the robot is to move beside the target cartridge to pick it up (GET). The \textit{final motion} (c2d) is to carry it to the drive (PUT) for read/write operation. The robots when being used stick to these four essential motions in the order presented.  The total exchange time is defined to be the sum of these four motions. The distributions of r2d, d2c, c2c and c2d are determined by the geometry of the library system, robot motion statistics and a mean value provided by the user. A sample distribution is provided in Fig. \ref{fig_sim2_dist} for a $36 \times 20$ 2D library in which $6$ drives are placed in the center with a robot operating at a mean exchange rate of  $250$xph. Note that for this exchange rate, the mean motion time is $3.6$ secs. 

\textit{Robot exchange count} is defined as the number of times the robot completes a full exchange motion as defined above. Robot exchange rate is the ratio of robot exchange count to total number of discrete simulation steps for a given period of time (e.g. an hour). In fact, maximum robot exchange rate is a wear parameter associated with each robot and usually expressed in terms of exchanges per hour (xph). From this metric, the minimum exchange time can be calculated to be used in our simulation platform. In large enterprise library systems for instance, there may be more than one robot. In order to balance the wear, we pseudorandomly choose one of the robots if more than one is housed in the corresponding PDRs. One of the performance metrics that is directly tied to exchange count is the \textit{number of objects touched (NoT)} in the tape cartridges. Everytime a robot moves from cartridge to an empty drive, we increment \textit{NoT} count. %In fact, this would be a lower bound on the number of objects touched if collocation is utilized. 

%\subsubsection{Object Size} It is reported in various past studies  that the data object size (stored or communicated as objects with an average size of $\mu_o$) tend to possess heavy-tailed distribution such as Pareto \citep{Satyanarayanan}. On the other hand, there are also studies demonstrating that other heavy-tail distributions might be possible for local file systems \citep{downey}, hard disk data \citep{arslan2020distribution} and world wide web \citep{gong}. In our study, we assume a Weibull distribution with configurable shape and scale parameters such that we can model fixed and exponential object sizes by adjusting these parameters.

%\subsubsection{Loading and Positioning} Media load time is defined as the amount of time between the cartridge insertion into a drive and the drive firmware become ready to listen for host system commands. We assume the media load time to exhibit only little variation as the drive system is an enclosed system and therefore least affected by external changes. After the cartridge is loaded into the drive, the head is being positioned on the location of the data laid on the magnetic tape. Depending on the exact data location on tape layout, the loading time may dramatically vary. We model the location of the data object to be uniformly distributed across the tape surface. 

\subsection{Advanced System Features}

{In this subsection, we will elaborate on the sophisticated functionalities of the library simulation model, including aspects such as collocation, data redundancy and protection support, and various protocol types employed during data retrieval from the library.}

%In this subsection, we shall detail the advanced features of the library simulation model which includes collocation, data redundancy and protection support, protocol types while retrieving data from the library. 

\subsubsection{Collocation} 

Collocation is a technique that involves collecting numerous small-sized data requests and assembling them into a larger chunk of request within a buffer. The objective of this technique is to decrease internal system traffic and alleviate queue load, which ultimately enhances access times. In other words, incoming data requests from a specific user are accumulated in the server RAM till the total size exceeds a \textit{threshold} value. If we let $m_i$ be the mean size of the requested data by user $i$, and $a_i = $\textit{threshold}$/m_i$ where \textit{threshold} is a common collocation parameter.  Then, the arrival rate of collocated requests of user $i$ will be another Poisson process with rate $\lambda/a_i$. Collocation between different users {is useful} only if their data content have some sections in common i.e., deduplication is {effective}. 

\subsubsection{Redundancy for Reliability}

Replication is a technique to maintain data reliability by storing multiple copies of the same exact data object in different failure domains. The number of copies is known as the \textit{replication factor} of the system. In the library model, different failure domains may mean different cartridges or different libraries depending on the single or multiple library (ML) contexts. For instance, if it is a single Enterprise library simulation it would mean different and independent cartridges assuming that cartridge wear or failures are relatively independent of each other.  On the other hand in a RAIL environment, it would mean different and independent libraries. Let us take the cartridge-level failure tolerance as an example. A user can request his data object without knowing the existence of the replicas. However, having multiple copies available can help reduce the access latency performance. There are two protocols our library system model supports to deal with the retrieval and post services. We detail these protocols in the next subsection.

On the other hand, replication leads to an inefficient storage resource utilization. In other words, the maximum usable raw data storage of the library system decreases  proportional to an increase in the replication factor. An obvious solution to this is the use of \textit{erasure coding} in which a data is first partitioned into $k$ equal size smaller objects (also known as fragments) and gets encoded into $n$ distinct objects of the same size. {Numerous open-source erasure coding libraries are available to facilitate the creation of redundancy for this purpose \cite{plank2008jerasure, arslan2021founsure}.} The group of all these $n$ objects is named as a codeword. These $n$ objects (also referred as the symbols of the codeword) are placed in the DR-queue for service just like before. In case of any $l \leq n-k$ failed cartridges or libraries, we would still be able to recover the data object from the remaining $n-l$  objects scattered across different failure zones. If $l=n-k$, the erasure codes are deemed to be storage-optimal and belong to the class of  Maximum Distance Seperable (MDS) codes \citep{wicker1999reed}. If $k$ raw data objects are part of the codeword, the erasure codes are called systematic. Otherwise, they are named non-systematic. Since data object reconstruction requires only $k$ fragments for MDS codes, one needs to put $s \geq k$ fragment requests on respective DR queue/s. The ways these fragments are requested are tied to a set of rules known as
\textit{protocols}. Depending on the protocol we use, the latency performance with erasure code redundancy may vary. 

\subsubsection{Redundant and Failure Protocols}

One of the most distinctive features of the proposed simulation model is the utilization of two different protocols under two different redundancy schemes, namely, redundant v.s. failure and replication v.s. erasure coding.  We shall detail these protocols for replication and erasure codes in separate subsections.

\paragraph{\ul{\textbf{Redundant Protocol with Replication}}} In this case, we place multiple copies inside the same library on different cartridges (failure zones). This protocol dispatches multiple calls of the same data (depending on the replication factor) and places these requests sequentially on the DR queue. 

\begin{figure}[!t]
\centering
\includegraphics[width=0.9\columnwidth]{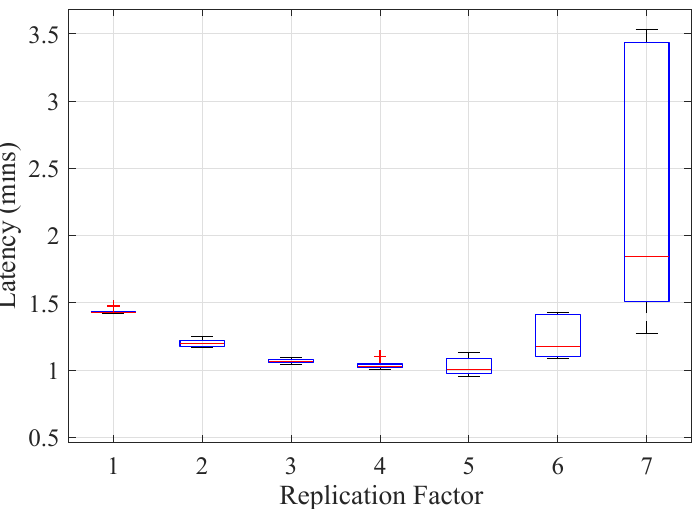}
\caption{A box-plot demonstrating the trade-off between latency and replication factor using \textit{Redundant protocol} (Geometry assumed: $25 \times 640$).}
\label{tradeoff_rep_lat}
\end{figure}

In various events of failure, e.g., cartridges may be unreadable due to aging or aggregated debris, and {successful reads by the} drives may require a lot of retires, the system dispatches as many copies as can be placed (not more than the replication factor) on the DR queue\footnote{We assume no catastrophic failures i.e., data comes back almost certainly (as long as an enough number of retries are used) though at the expense of increased latency. In other words, dead cartridge and permanent data inaccessibility is ignored throughout this study.}.  These copies share the same timestamp but different message IDs for later processing. The system does not wait for any copy request to successfully return to take any further action. In other words, the system operation is completely asynchronous. The earliest copy service completion (arrival) shall be served to the user who initiated the request. The disadvantage of replication is two fold. First, it consumes a lot of storage space in the library resulting in inefficient utilization of cartridge capacity. Secondly, due to multiple dispatches, the rest of the replies are simply ignored and will create internal bus/link and queue traffic. Asynchronous multiple dispatches can help with the latency performance of the system. This is due to libraries {having} multiple robots i.e., enterprise libraries, drives and other resources that operate independent of each other. However, both library queues are inherently sequential. We can place two requests of the same data one after another (with different IDs) and the library system will allocate resources for these copies in a first come first serve fashion.

Indeed, there is a trade-off between the replication factor and access latency as shown in an example run for $25 \times 640$ geometry in Fig. \ref{tradeoff_rep_lat}. As can be observed, having more independent copies stored in different cartridges help improve the latency performance till we reach up to 4 copies for this particular geometry. Beyond that, latency performance is adversely effected due to increased traffic inside the library queues. {Furthermore, increased traffic leads to instability in the queues resulting in more variability in the latency performance, particularly beyond the redundancy more than sufficient. In the illustrated example depicted in Fig. \ref{tradeoff_rep_lat}, it is observed that employing four copies results in optimal latency performance. Beyond this threshold, not only does the mean latency degrade, but also its variance increases.} This simple demonstration also advocates that using the proposed simulation model, access-latency-optimal number of copies can be determined for a given set of system parameters. 

\paragraph{\ul{\textbf{Failure Protocol with Replication}}} In case of failures, the system dispatches only one copy request (randomly\footnote{In fact closest copy inside the library based on the geometry might be preferable in terms of data access latency performance. However, this would require extra logic implemented in the server as well as a database (or a computation framework) to access the locations of the stored content.}) and places it on the DR queue. The system attempts to bring the data object back while it blocks other requests for the same data. If the placed data request never comes back within a certain time threshold, other copy requests are subsequently placed in the queue until the user’s request is succesfully served. If none of the copy requests return within the allotted threshold time, the data is considered lost or temporarily unavailable which results in system halt.

One of the reasons for a failed response of requested data may be due to a failure that drives attempts to solve with repeated re-tries. In case the drive is unable to read the data object (which happens with some non-zero probability), it is capable of retrying the data read sessions by repositioning the drive heads. We assume that the number of retries follows a binomial distribution. Also, since the drive systems are enclosed stable devices operating at around 300MBs/sec.\footnote{This parameter is also configurable for the future generation of LTO drives.}, each re-try is assumed to take almost the same time. We do not allow each copy service time to take forever by thresholding the maximum wait time. If the re-positioning and read operations do not terminate successfully within that time period, the next copy request is placed on the DR-queue. In case of success, we no longer place any more data requests on the queue and help create less congestion on the DR and R queues and eventually obtain better access latency performance.

\paragraph{\ul{\textbf{Protocols with Erasure Coding}}} 
The selection of erasure coding as the redundancy scheme represents a generalization of the replication approach, where $k=1$ represents a single replication and $n$ denotes the replication factor. However, if the condition $n>k>1$ is met, we specifically employ the term "erasure coding" to describe this particular redundancy scheme.

\textit{Redundant protocol} dispatches a subset of size $s \ (\geq k)$ of the data fragments and places these requests sequentially on the DR queue to be served by robot/s. These fragment requests share the same timestamp and message ID. The server system waits until earliest $k$ of these $s$ fragments are serviced successfully. If the number of returned fragment requests is less than $k$, then the system throws an “unable to retrieve the data object” error. Otherwise, the requested object is reconstructed using the collected $k$ fragments by a subsequent decoding operation. 

In order to describe the \textit{Failure protocol}, let us introduce some notation. Let $\mathcal{N}_n = \{1,2,\dots,n\}$ to represent {the ordered set of} fragment indexes. Failure protocol dispatches only a $k$-size subset of the data fragments $S_k \in \mathcal{N}_n$ and awaits for a response. If all of these requests return successfully, the data is served to the user. Suppose $t$ fragment requests never arrive (i.e.,  the service time exceeds a given time threshold for these fragment requests). In that case, the system places any $t$ of the fragments from the set $\mathcal{N}_n - S_k$ on the DR queue. Note that we implicitly assume $t \leq n-k$. Otherwise, the system throws “unable to retrieve the data object” error. If system uses systematic MDS code, then $k$ fragments are data fragments themselves and only in case of failure, decoder is run. On the other hand, if system uses a non-systematic MDS code, the decoder needs to run all the time which shall prolong the total service time. Note that if there is no failure, Failure protocol leads to less congestion on the DR and D queues. 

\begin{figure}[!t]
\centering
\includegraphics[width=\columnwidth]{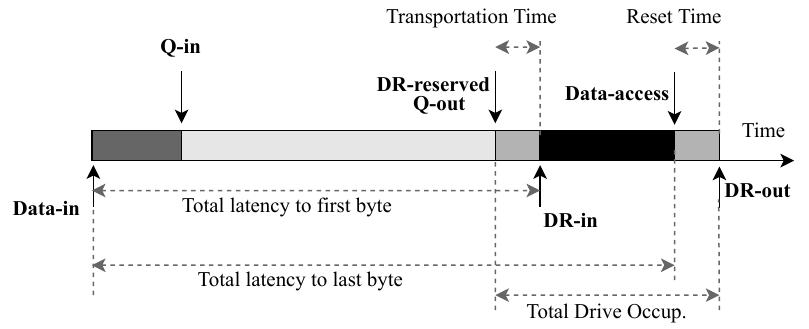}
\caption{The operation cycle of a data retrieval request.}
\label{fig_life_cycle_data}
\end{figure}

\subsubsection{The operation cycle of a data retrieval request}

In Fig. \ref{fig_life_cycle_data}, we demonstrate several check points in the operation cycle of a data request. Data-in represents the time instant where a particular data request meets the library system for the first time. Q-in denotes the time when that request is put in the DR queue. Whenever a drive and a robot are reserved, the request is pulled out of the DR queue at Q-out and after some transportation time, the cartridge is inserted into the drive at DR-in and the request receives service. The transportation time includes c2c, d2c, c2d. The drive is loaded with the cartridge and the tape is rolled to the exact location where the data is stored. After reading data, the retrieval service terminates at Data-access. Thus, the sum of loading and read times is given by the difference between Data-access and DR-in. Right after the completion of the requested task, the drive is reset and becomes available again to serve the next request. This reset time includes repositioning, unloading and r2d. Similarly, the total latency to the first and the last bytes are given by DR-in $-$ Data-in and Data-access $-$ Data-in, respectively. The total drive occupation time is also given by DR-out $-$ Q-out. With this checkpoint definitions, various other performance metrics can be defined in similar fashion. 

We also have to note that Data-in and Q-in need not be identical depending on the choice of the protocol we use. In the failure protocol for instance, the first copy request may never go through due to read failures and hence a second request might be inserted into the DR queue for the same exact data. In this case, the time Q-in characterizes the second request being placed in the DR queue which is not necessarily equal to Data-in time for the same data request.

\subsubsection{Rate of Object Request} The data object request rate can be determined either by manually setting $\lambda$ to an appropriate value or based on the {amount of} raw data stored on the library system. Given the parameters of the system and time period $T$, we can compute the rate of the Poisson arrival rate within $T$ as
\begin{eqnarray}
\centering
\lambda = \frac{NoC \times C_t \times \Phi_f \times \textrm{AOTR} \times k}{n \times \mu_o \times T} \label{rateofreq}
\end{eqnarray}
where  $\Phi_f$ is the library fill ratio. The aggregate rate of request needs to be modified accordingly as the simulation settings change. For instance in a RAIL-type ML simulation setting under different protocols and the number of users, aggregate rate of request arrivals are distributed among different independent libraries reducing the rate of request rate per library. 

\section{Multiple User/Library Simulations}

%A summary of the ML system is depicted in Fig. \ref{fig_mlx}. A central server performs multiple jobs such as dispatching and load balancing. The load balancer duplicates and places appropriate requests (a total of at least $k$) into the DR queues of multiple independent tape libraries. 

%A depiction of the ML system can be seen in Fig. \ref{fig_mlx}. The ML system comprises a central server that performs several tasks, including dispatching and load balancing. The load balancer duplicates and allocates suitable requests, totaling at least  $k$, across the DR queues of multiple independent tape libraries.

\begin{figure}[!t]
\centering
\includegraphics[width=0.95\columnwidth]{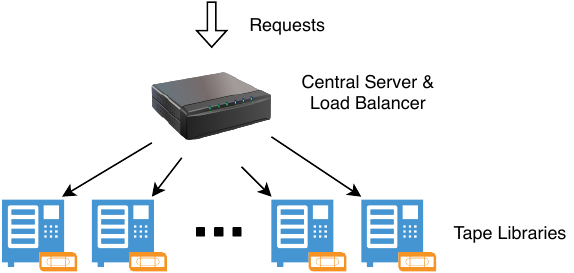}
\caption{A centralized ML system architecture.}
\label{fig_mlx}
\end{figure}

A simplified illustration of the ML system is presented in Figure \ref{fig_mlx}. This system consists of a central server responsible for various functions, including dispatching and load balancing. The load balancer duplicates and assigns appropriate requests, totaling a minimum of $k$, among the DR queues found in numerous distinct tape libraries.

In ML, individual libraries maintain their distinct queues. Upon the arrival of a data object request, we simultaneously direct the identical request to a minimum of $k$ Data {Request} (DR) queues. Homogeneity is assumed across libraries, sharing identical parameter lists. Leveraging this assumption, we implement code reusability to emulate concurrency by employing the software designed for a singular Enterprise library\footnote{By playing with the configuration parameters, these independent libraries become no longer Enterprise.}. 
We extend our assumption that every autonomous library encounters identical data request patterns. To replicate the data request arrivals across distinct libraries, we employ selective seeding for the random number generators. Despite the sequential operation of each library, the resulting output files emulate multiple concurrent libraries handling indistinguishable data object request patterns. This is facilitated by selective seeding, ensuring that the random number generators generate precisely matching sequences for each independent library execution. 
%We further assume that each independent library confronts with the same data request patterns. To regenerate the data request arrivals for each distinct library, we apply selective seeding for the random number generators. Although each library will run sequentially,  the output files shall be generated as if we run multiple concurrent libraries loaded with identical data object request patterns. This is possible due to selective seeding which makes sure that the random number generators create the exact same sequences for each independent library run.  

%In case of ML, contrary to single Enterprise, both Redundant and Failure protocols place the data object (fragment) requests on different DR-queues. This is due to libraries are assumed to be in different failure zones. This assumption is quite reasonable because otherwise generation of more redundancy will be ineffective.  
Unlike the single Enterprise scenario, both Redundant and Failure protocols allocate data object (fragment) requests to distinct DR-queues. This divergence is attributed to the presumption that libraries exist within separate failure zones. This assumption is deemed rational as generating additional redundancy would be futile otherwise.With respect to Redundant protocol, similar to single enterprise library scenario the earliest $k$ data service arrivals shall be used to decode and serve the user data store/retrieval request. In the Failure protocol however, when a request is delayed, an appropriate selection of a library is made to serve in place of the delayed request. However, this would lead to dependent simulation of independent libraries. In other words, failures in one library (due to absence of redundancy within a library) will lead to a change on the request patterns of another library. The way we resolve this issue and allow independent simulations of multiple libraries is to use averaging arguments. In other words, the additional data object requests due to potential failures in other libraries can stochastically be modeled into a random variable and the data request patterns can accordingly be adjusted to compensate for these additional requests due to potential failures. Additionally, library systems typically serve more than one user and since the load balancer dispatches user requests in a pseudo random fashion, the total number of requests in a library DR-queue would become random.

Suppose we have $L$ users and $N$ library nodes, and the $i$th user has {an annual object touch rate of} $\phi_i$. Let $m_i$ be the random variable characterizing the number of user fragment requests for library node $i$. We assume $m_i$ to be binomially distributed, {expressed as,}
\begin{eqnarray}
{m_i \sim } \ p(L, {x}) = \binom{L}{x} \left(\frac{s}{N}\right)^x \left(1 - \frac{s}{N} \right)^{L-x}
\end{eqnarray}
for $x=1,2,\dots,L$ {and $s/N$ characterizes the probability of using the $i$-th library fragment since only $s \geq k$ data fragments are placed in the queue as requests}. Let us suppose that $d_{i,j}$ represents the data access (to first or last byte {depending on the context}) latency of the $i$-th library for the $j$-th data object retrieval and $m_i$ to be the total number of requests in the $i$-th library DR-queue. %Also, denote $\min^{(l)}_i m_i$ to be the $l$-th minimum of all $m_i$s. 
The overall data access latency of the ML system is {therefore} approximated by $
\approx \frac{1}{\min m_i} \sum_i {\min_j}^{(k)} d_{ij}$
where $``{\min_j}^{(k)}"$ {operator finds} the $k$-th minimum. Here, the fragments that belong to the same codeword stored on different libraries are identifed based on their message IDs.

While parallel processing enhances the service quality of RAIL-type Machine Learning (ML) systems, the component libraries are generally chosen to be less powerful and cost-effective. Conversely, individual libraries tend to be at an enterprise level, furnished with multiple high-quality robots and drives. A structural contrast between RAIL-type ML setups and single Enterprise libraries lies in their discrete queues, despite both systems potentially featuring an equivalent (or comparable) total count of drives and robots. This distinct queuing system enables ML to efficiently handle requests in a highly parallel manner, thereby enhancing latency performance.

%Although parallel processing would enable RAIL-type ML systems to demonstrate better service quality, the constituent libraries are typically selected to be less capable and economical (inexpensive). On the other hand, single libraries are usually enterprise-level and equipped with  multiple high quality robots and drives. One of the structural differences of RAIL-type ML case compared to single Enterprise library is the distinct queues although both systems may have the same (or similar) number of drives and robots overall. This allows ML to process requests in a highly parallel fashion resulting in increased latency performance. 

In Failure protocol, we take a probabilistic approach for RAIL-type ML simulations. First off, the rate of the arrival for each library node is given by $\lambda_j = s \lambda/ N$. However since some requests will return error due to wait times are larger than the predetermined threshold, central server will dispatch new requests to other distinct libraries. This will increase the total request rate on each library  to $\lambda_j^\prime > \lambda_j$. In fact, in each read failure, AOTR would be increased by a factor of $(n-k)(N-1)/N$. In other words, with the drive failure probability $p_d$, we can calculate $\lambda_j^\prime$ using equation \eqref{rateofreq} based on averaging arguments. In that case, the updated average annual object touch rate would be given by $\phi_i = \frac{(n-k)(N-1)}{p_dN}$.

\section{Read Latency Analysis}

%The basic tool to use for the analysis is based on Queuing theory \citep{7774566}. First, we realize that although the arrivals are Poisson, service time is a function of library geometry and is nowhere near one of the familiar distributions. There are two queues DR and D which are highly dependent on the availability of the library resources. Hence queuing times as well as the services times are directly affected by each other. Clearly, service time is not exponentially distributed.  On the other hand, assuming each drive retry takes exponential time, and sum of exponentially distributed retries shall give us a Gamma distribution.

The primary analytical method employed here relies on Queuing theory \citep{7774566}. While arrivals follow a Poisson distribution, service time is intricately linked to library geometry and doesn't align with commonly known distributions. The system comprises two queues, DR and D, both heavily reliant on library resource availability, which directly influences queuing and service times. It's evident that service time doesn't adhere to an exponential distribution. However, considering each drive retry follows an exponential pattern, the summation of exponentially distributed retries yields a Gamma distribution.

%Let us remember Kendall's notation for the queuing systems. We refer to $T/X/C/K/P/Z$ queuing system where $T$ is probability distribution of inter-arrival times, $X$ is probability distribution of service times, $C$ is number of servers, $K$ is queue capacity, $P$ is the size of the population and $Z$ is the service discipline.

%To recall Kendall's notation for queuing systems, we use the notation $T/X/C/K/P/Z$. This refers to a queuing system where $T$ represents the probability distribution of inter-arrival times, $X$ represents the probability distribution of service times, $C$ represents the number of servers, $K$ represents the queue capacity, $P$ represents the size of the population, and $Z$ represents the service discipline. For instance $M/M/1$ would characterize Poisson request arrivals, exponential service times and there is only single robot/drive as a server. In our approximate analysis, we typically assume large size queues. Thus, our queues in the library system can be assumed to act as a $M/G/k/\infty$ (or $M/G/k$ in short) system where $k$ represents the number robot/drives. Unfortunately, most performance metrics for this queuing system are not known. Plus, our model consists of two-phase dependent queues (DR and D) and use of two different protocols which would complicate the overall operation model. Thus, there seems to exist no easy mathematical resolution for the proposed system.

Let us recall Kendall's notation for queuing systems expressed in the form {$\mathcal{T}/\mathcal{X}/\mathcal{C}/\mathcal{K}/\mathcal{P}/\mathcal{Z}$, where $\mathcal{T}$} refers to the probability distribution of inter-arrival times, {$\mathcal{X}$} denotes the probability distribution of service times, {$\mathcal{C}$} stands for the number of servers, {$\mathcal{K}$} signifies the queue capacity, {$\mathcal{P}$} indicates the size of the population, and {$\mathcal{Z}$} represents the service discipline. For instance, $M/M/1$ characterizes Poisson request arrivals, exponential service times, and a single robot/drive serving. In our approximate analysis, we typically assume large queues. Therefore, our library system's queues can be approximated as an $M/G/c/\infty$ (or simply $M/G/c$), with $c$ representing the number of robot/drives. Unfortunately, most performance metrics for this queuing system remain unknown. Additionally, our model comprises two-phase dependent queues (DR and D) and employs two distinct protocols, which complicates the overall operational model and its analysis. Consequently, there appears to be no straightforward mathematical resolution for this proposed system.

Let's direct our attention to a single queue (or isolated queues) to simplify {our} analysis. Disregarding the interplay between the queues (treating them as separate entities) and employing a redundant protocol in conjunction with $c$ number  of robots/drives, an approximation for the mean number of requests awaiting service in the queues can be derived as follows \citep{Ross2014}
\begin{eqnarray}
L_q \approx \frac{P_0 \rho^{c+1}}{c! (1-\rho)^2}
\end{eqnarray}
where $\rho = \frac{\lambda}{c\mu}$ and $P_0$ is the probability that there are no requests waiting in the queue and it can be approximated by
{
\begin{eqnarray}
P_0 \approx  \left[\sum_{m=0}^{c-1} \frac{(c\rho)^m}{m!} + \frac{(c\rho)^c}{c!(1-\rho)}\right]^{-1}
\end{eqnarray}
which follows from standard analysis for $M/M/c$ queuing model for multi servers.}

\begin{figure}[!t]
\centering
\includegraphics[width=\columnwidth]{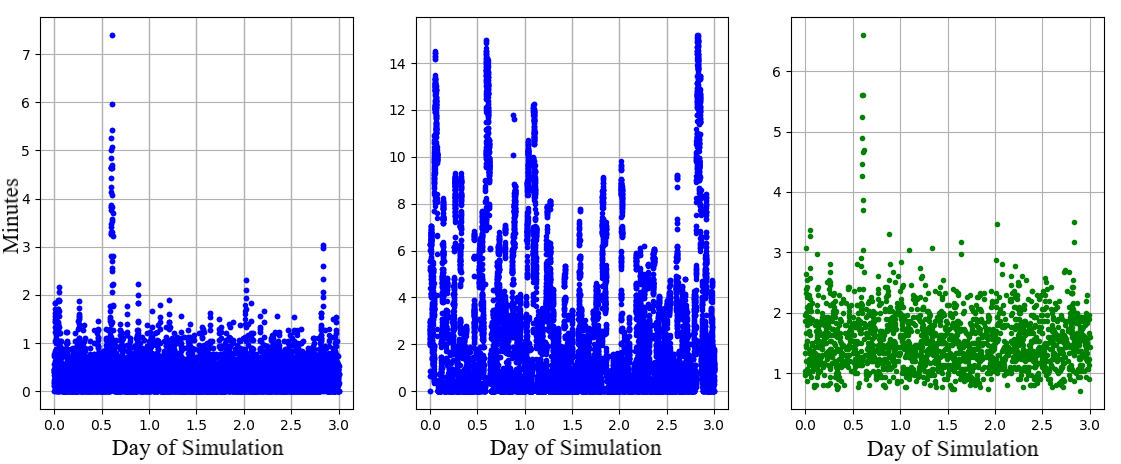}
\caption{DR and D queue performances and data access latency (all in same units) as a function of days passed (Redundant Protocol).}
\label{fig_ml}
\end{figure}

\begin{figure}[!t]
\centering
\includegraphics[width=\columnwidth]{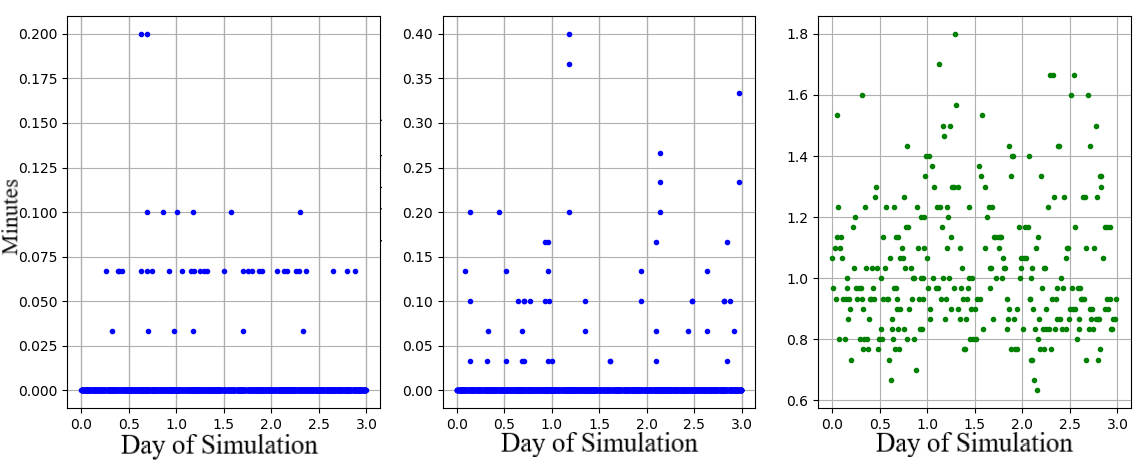}
\caption{DR and D queue performances and data access latency (all in same units) as a function of days passed (Failure Protocol).}
\label{fig_ml2}
\end{figure}

Finally, using Little's law, we can find the wait time in the queue to be approximated by $W_q \approx L_q / \lambda$. On the other hand, to find $W_q$ for a more general $G/G/c$ queue model, we can use the following approximation \citep{Ross2014}
\begin{eqnarray}
G_q \approx W_q \frac{C_a^2+C_s^2}{2}
\end{eqnarray}
 where $C_a^2$ and $C_s^2$ are the coefficients of variation of the inter-arrival and service times, respectively. In case of exponential inter-arrival and service times, we have $G_q \approx W_q$. To be able to find the data access time, we focus on the DR queue. However, for continued service we would need both a drive and a robot.

For a very rough approximation, we decouple the service time of Robot+Drive by introducing a fictitious queue (queue \textit{B}) and combine two separate robot services in another fictitious queue (queue \textit{A}) to model the overall behaviour. In fact, assuming $r$ robots, and $d$ drives, the queue \textit{A} can be modeled as $M/G/r$ whereas queue \textit{B} can be modeled as $G/G/d$ due to arrivals would no longer be according to Poisson. Assuming average robot and drive service times to be $s_R$ and $s_D$, the average access time to data would be given by
\begin{align}
    W_q^{({A})}+W_q^{({B})}+s_R+s_D.
\end{align}
The duration of robot and drive servicing relies on various factors, including the robot-to-drive and drive-to-robot distributions, geometric configurations, head positioning, load time distributions, and the employed protocol. It's worth highlighting that our derived expressions serve as upper (or lower depending on the context) limits for assessing the comprehensive performance of library systems, representing idealized scenarios.

\begin{figure}[!t]
\centering
\includegraphics[width=\columnwidth]{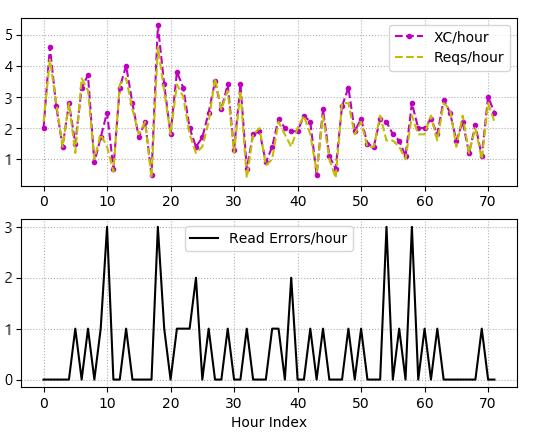}
\caption{Total read errors/exchanges/requests as a function of time (in hours). XC/hour: exchanges per hour.}
\label{fig_xo}
\end{figure}

\begin{figure*}[!t]
\centering
\includegraphics[width=2\columnwidth]{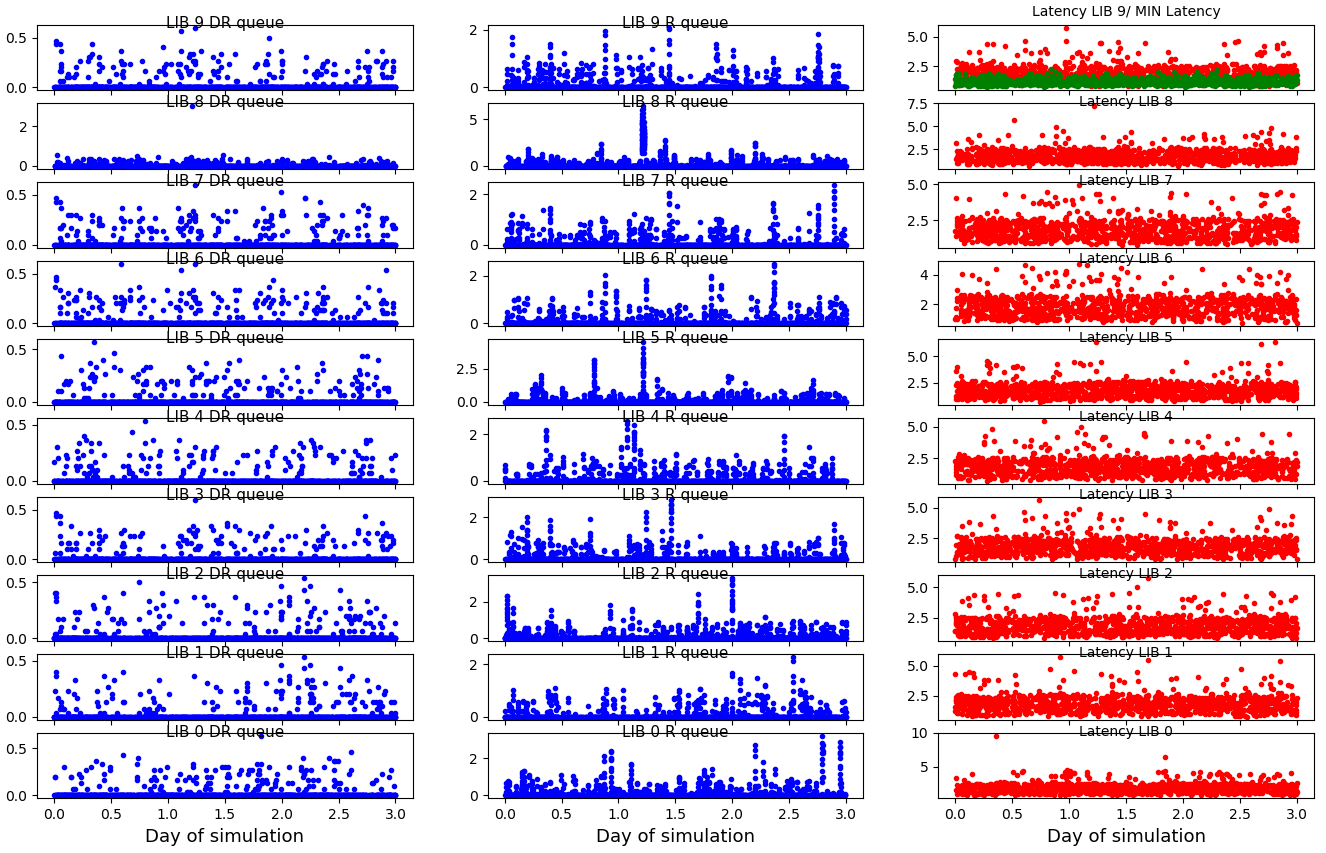}
\caption{DR and D queue performances and data access latency (all in same units) as a function of days passed for ten parallel libraries (Redundant Protocol). Top right corner also includes the plot of minimum data access latency after the data retrieval. {It is not hard to observe the significant improvement due to using multiple libraries with Redundant Protocol.}}
\label{fig_ml3}
\end{figure*}

%Note that robot and drive service times are functions of robot-to-drive, drive-to-robot distributions, geometry, head positioning, load time distributions and the protocol we use. We finally note that our expressions can used to upper bound the overall performance of library systems as they describe idealized scenarios.

\section{Numerical Results {\& Discussions}}

%In this section, we provide some representative simulation results of our tape simulation platform. Throughout this section, we assume no collocation in order to focus on the worst case resource utilization performance and report unitless and/or relative results to illustrate statistical and comparative rather than raw performances.  

In this section, we present selected simulation {results} from our tape simulation platform. We assume no collocation throughout this section to emphasize the resource utilization performance under the worst-case scenario. The results reported are unitless and/or {normalized} to demonstrate statistical and comparative performance instead of raw performance.

We begin by simulating a single Enterprise library system containing $40 \times 168$ (a total of $6720$) cartridges each with capacity 12TBs, two high-quality robots (operating at 150xph each) to serve both queues and 80 independent drives each with an average of 300MB/s streaming data rate\footnote{Note that this number is arbitrarily chosen. One could adapt for LTO8 drives (released in 2017) which operates at 360MB/s, and LTO9 drives (released in 2021) operates at 400MB/s.}. {For all our simulations,} we assume 18 secs and 50 secs average drive load and positioning times (based on typical TLO6 library performances), respectively, each one uniformly distributed with support starting at 0. The data object size is assumed to be fixed at 5GB (non-random). We used $(n=6,k=1)$ erasure code, i.e., 6-copy system with both \textit{Redundant} and \textit{Failure} protocols using 600 objects touched on average per day (which is manually set) by 40 distinct users. The Failure protocol would be using 100 simulation steps as the decision threshold unless otherwise stated. Assuming $p_d = 1\%$ drive read failures occur\footnote{This rate might not be too realistic, however is set such to clearly illustrate the points we would like to cover with our simulation platform.}, with a maximum of 10 drive re-tries, we have simulated the single tape library system for 3 days time period (72 hours) and obtained the average queue lengths as well as the data access latency to the last byte as a function of time (in hourly discrete steps).

%In Fig. \ref{fig_ml} and Fig. \ref{fig_ml2}, DR and D queue waiting times as well as data access latency are shown all in terms of the same unit of time as a function of day time in hours for Redundant and Failure protocols, respectively. Accordingly, with the given parameter selections, the simulation result indicates that it takes on average \% 48 more time to retrieve data from the library system using Redundant protocol relative to Failure protocol. As expected, Failure protocol also touches objects a little more than one-sixth of the total number of objects touched using Redundant protocol. This feature allows better utilization of the library resources and hence better latency performance at the expense of more complex implementation and more data access. 

In Figs \ref{fig_ml} and \ref{fig_ml2}, the waiting times for DR and D queues, along with data access latency, are depicted in terms of the same time unit, in terms of the daytime hours for Redundant and Failure protocols, respectively. Based on the specified parameter configurations, the simulation results reveal that, on average, it requires \% 48 more time to retrieve data from the library system using the Redundant protocol compared to the Failure protocol. As anticipated, the Failure protocol also interacts with objects, slightly exceeding one-sixth of the total number of objects affected by the Redundant protocol. This characteristic enables more efficient utilization of library resources, resulting in enhanced latency performance, albeit at the cost of a more complex implementation and increased data access.

%In Fig. \ref{fig_xo}, one of the other simulation outputs is shown which demonstrates when and how many read errors have happened during the single library simulation while using Failure protocol. In our context, a read error happens when the drive executes all possible re-tries (10 for this simulation) and fail to retrieve the data within a threshold time (100 steps for this simulation). As a consequence of this, by playing with the threshold parameter the read errors can be increased or decreased at the expense of better or worse access latency performance. 

In Figure \ref{fig_xo}, an alternative simulation output is presented, illustrating the occurrences of read errors during a single library simulation when utilizing the Failure protocol. In our specific context, a read error is defined as an event where the drive exhausts all permissible re-tries (set at 10 for this simulation) and fails to retrieve the data within a predefined threshold time (100 {discrete time} steps in this simulation). Consequently, manipulation of the threshold parameter allows for the intentional increase or decrease of read errors, impacting access latency performance positively or negatively.

Moreover, the graph overlays the total number of exchanges per hour (XC/hour) and incoming requests per hour (Reqs/hour), revealing a clear proportional relationship between the load on robots and incoming data retrieval requests. This correlation is primarily attributed to the utilization of the Failure protocol and the relatively low drive read failure rates. Had the Redundant protocol been employed instead, the system would have generated three times as many requests, leading to a corresponding increase in robot exchanges. Such a graphical representation is crucial for monitoring the system's health. Thanks to the multiple copies inherent in the Redundant protocol, the library retains the capability to respond to diverse data object requests initiated by users.

%In addition, the total number of exchanges per hour (XC/hour) as well as incoming requests per hour (Reqs/hour) are also plotted on top of each other which clearly demonstrates that the load on robots is proportional to the incoming data retrieval requests. This is mainly due to the use of Failure protocol and relatively small drive read failure rates. If we had used Redundant protocol instead, the system would have generated three times more requests and accordingly increased robot exchanges. Such a plot might be important for monitoring the health of the system and thanks to the multiple copies, the library would still be able to respond to the various data object requests originated by the users. 

\begin{figure}[!t]
\centering
\includegraphics[width=\columnwidth]{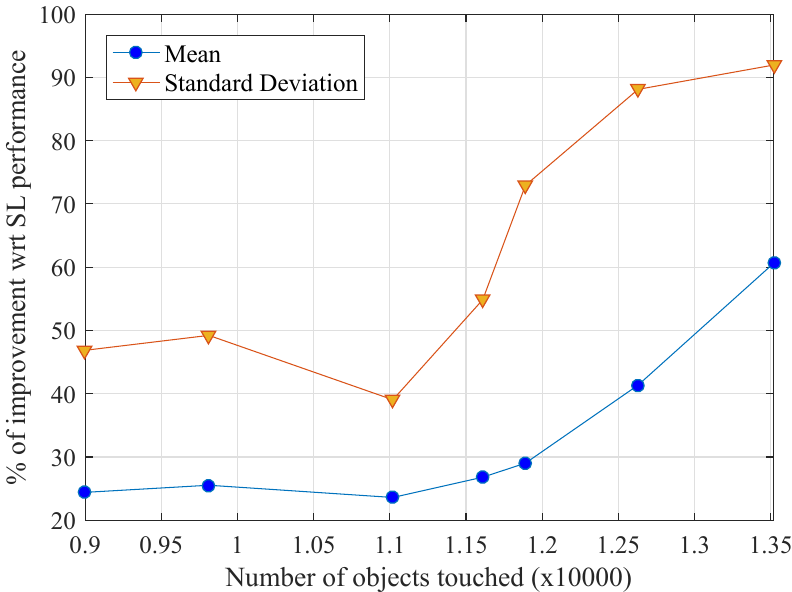}
\caption{Relative increase in the average data access latency performance with respect to (wrt) single enterprise library using RAIL-type scale-out libraries as a function of number of objects touched.}
\label{fig_slvsml}
\end{figure}

In order to compare a scale-up Enterprise  and a scale-out RAIL-type  library configurations, we also simulated a multiple-library distributed tape system. In this case, the component libraries are selected to be of a geometry $21 \times 32$ (a total of 672) each cartridge with 12TBs capacity. To be able to fix the system size and the number of drives, we have used 10 libraries, each is equipped with 1 low quality robot (operating at 100xph) and 8 same-quality drives. Note that with this configuration, the library size is set to be 80.64TB which is identical to that of the single Enterprise case. The rest of the simulation parameters are assumed to be the same.  The results are plotted in Fig. \ref{fig_ml3}. We realize from the results that with the RAIL-type scale-out setting the queue loads are considerably decreased providing a mean latency improvement of 25\% of relative to single enterprise library. We also note that both systems use Redundant protocol and similar number of objects are touched, which is mainly  characterized by 600 objects touched on average per day. This parameter of the system determines the average traffic that is imposed upon the library system/s. In order to clearly demonstrate the main benefit of the RAIL-type ML system, we gradually varied this parameter to change the total actual number of objects touched in our 3-day simulation. We run several simulations with different traffic conditions and plotted the results for both Enterprise and RAIL-type ML cases in Fig. \ref{fig_slvsml}. As can be observed from the plotted results in a relative manner. In other words, we have plotted $\%$ improvement of single Enterprise library mean access time and its standard deviation as compared to the performance of RAIL-type multi library. Clearly evident from the results is the fact that the ML approach not only enhances overall average latency performance but also contributes to reducing the standard deviation of data access latency. The findings further indicate that, once the number of touched objects surpasses approximately 11500, there is a discernible acceleration in the growth of average access time and standard deviation for a single Enterprise library. This phenomenon is attributed to the library becoming unstable in response to the heightened rate of incoming requests. %As can be clearly seen, ML approach not only helps us get better average latency performance but also helps with the standard deviation of the data access latency. Results also show that  as the number of objects touched goes beyond around 11500, the average access time/standard deviation of single Enterprise library begins to grow more rapidly. This is seen to be due to the library going unstable due to increased request rate. 

\begin{figure}[!t]
\centering
\includegraphics[width=\columnwidth]{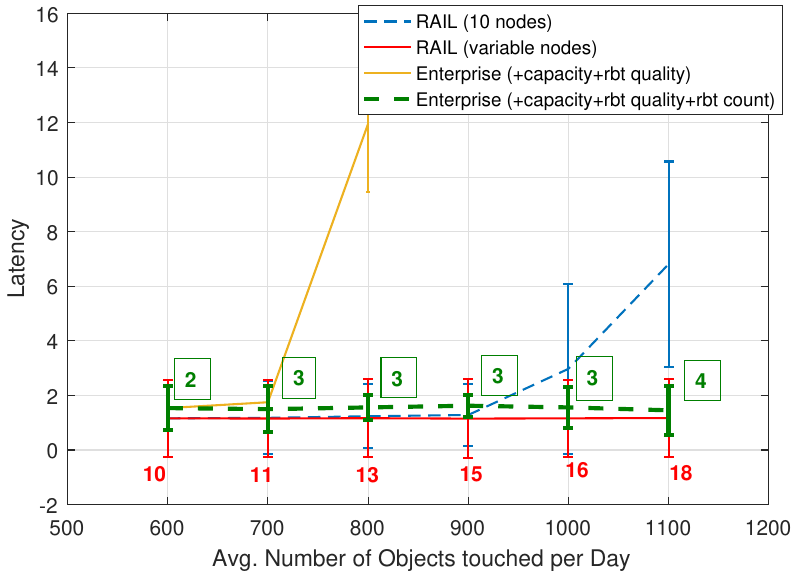}
\caption{Latency (mins) as a function of number of objects touched for RAIL-type scale-out system with fixed and variable number of nodes. In the variable case, the number of robots used are indicated for each data point.}
\label{fig_10vsvar}
\end{figure}

In Figure \ref{fig_10vsvar}, we illustrate the latency performance and the efficacy of the RAIL-type setting under escalating system demands. This demand is quantified by the manually set number of objects touched. In one scenario, we maintain a fixed number of nodes at 10, employing a 6-copy redundancy across all schemes. With an escalating demand, the system fails to generate responses promptly, resulting in degraded latency performance. Conversely, by allowing the number of nodes to incrementally increase, such that each node handles 60 object touches, we establish the RAIL system with variable nodes. Specifically, the number of nodes is set to $\lfloor$number of objects touched/60$\rfloor$. As depicted, the scale-out RAIL system adeptly accommodates the heightened demand, effectively maintaining latency performance at an acceptable level.

%In Fig. \ref{fig_10vsvar}, we have presented the latency performance and the potential of the RAIL-type setting with increasing demand from the system. This demand is quantified in terms of number of objects touched set manually. In one case, we assumed the number of nodes to be fixed to 10 and use 6-copy for redundancy in all schemes. As the demand increases, the system does not adaquately generate response on time and hence the latency performance degrades. However, if we let the number of nodes increase gradually such that 60 objects touches per node is observed, we obtained the RAIL system with variable nodes. In other words, the number of nodes is set to $\lfloor$number of objects touched/60$\rfloor$. As can be observed, the scale-out RAIL system addresses the increased demand and keeps the latency performance at an acceptable level. 

To be able to compare this adaptive RAIL system against the enterprise, we first considered a single scale-up system with the number of cartridges, drives and the quality of constituent robots (2 robots) in terms of xph scale all together as the number of nodes increase in order to maintain the same level of capacity and service quality. What we have observed with this setting is the unstable system operation after 700 object touches on average per day. {To mitigate this challenge, it was discerned that the scale-up system necessitates the inclusion of additional robots.} Those numbers are provided within boxes for each data point in Fig. \ref{fig_10vsvar}. Even with this adaptation, it is evident that the RAIL system demonstrates superior performance compared to the Enterprise system. This is primarily attributed to the independent queues, which adeptly handle incoming requests in a more flexible and efficient manner, utilizing lower-cost robots and drives—thereby operating at a reduced cost state. %As can be seen even with this adaptation RAIL system exhibits better performance compared to the Enterprise due mainly to independent queues serving the incoming independent requests more flexibly and efficiently with reduced quality robots and drives i.e., at a reduced state of cost.

\vspace{-3mm}
{
\section{Limitations and Challenges}}

%{One of the limitations of the proposed simulator is its 2D geometry. However, overcoming this limitation should not be too hard as the topological structure of the modern tape library systems can be modelled in 3D and be integrated into the simulator platform. However the current implementation does not allow a systematic fix such as modification of a configuration file. The required changes are still made into the source code directly to take effect. Another limitation of the present work is the simple data request assumptions which may not model the real-world complex workloads. Maybe more realistic data workload modelling can be performed and integrated into the simulator platform. Additionally, the deferred dismount is currently automatically determined depending on the availability of robot and drives. Alternatively, this variable can be implemented to be an independent parameter of the simulator that can be set externally for further optimizations. }

%One of the constraints of the proposed simulator lies in its 2D geometric representation of the library topology. Nonetheless, addressing this constraint should be feasible given the ease to model the topological structure of modern tape library systems in three dimensions and integrate it into the simulator platform. However, the present implementation lacks a systematic method for such integrations, such as modifying a configuration file. Instead, required adjustments are ought to be directly applied to the source code for implementation. 

{One of the physical limitations of the proposed simulator is its reliance on a 2D geometric (planar) representation to characterize the library's topology. This topological structure imposes limitations on the number of trajectories available for robots to facilitate the collaboration between the drives and cartridges. Despite this constraint, overcoming it appears feasible, particularly given the known nature of modeling contemporary tape library systems' topological complexities in three dimensional space. In other words, integrating more realistic topological enhancements into the simulator's framework seems appealing and realizable easily with the proper system configuration. Nevertheless, the existing implementation lacks a systematic approach for embracing these integrations, such as through the modification of a configuration file. Rather than employing a streamlined method, with the current version, necessary adjustments must be manually applied directly to the source code, which represents a less efficient and more labor-intensive process. }

Another {outstanding} limitation of this study pertains to the simplifying assumptions regarding data requests, which may not adequately capture the complexities of real-world workloads. Enhanced realism in workload modeling could be achieved and subsequently integrated into the simulator platform. {These workloads should be formulated based on potential application scenarios, and the performance of the library for each scenario should be assessed to determine its adherence to the required criteria.} Furthermore, the determination of deferred dismount currently relies on automated assessment based on robot and drive availability through queuing models. Alternatively, deferred dismount process could be implemented as an independent component of the system within the simulator {without direct dependency with queuing,} allowing for external configuration to facilitate further optimizations.

%{In the implementation phase, the multi-library simulation could either be run using independent parallel treads which would require the avialability of multi-processor envioronments. To make the simulator runnable on single processor environments, the implementation has faced to challenge of running multiple library systems sequentially simulating the same time process multiple times for different library systems. This has required a careful initilization of random number generators and simulating the across-library interactions.}

In the implementation phase, the multi-library simulation {could have been} executed using independent parallel threads, necessitating the support of parallel hardware and the availability of multi-processor runtime environments. Adapting the simulator to run on single-processor environments, while still being able to generate outcomes identical to that of purely concurrent processing, presented a challenge. The current implementation addressed this dynamic by sequentially running multiple library systems with identical {series of pseudorandomly generated} time stamps, allowing the simulation of the same time process repeatedly for each system. This required {careful} initialization of random number generators and {sequential} simulation of individual library-specific interactions along the common time axis. However, the inability to simulate advanced features, such as dynamically adapting and using the output of one library as the input for another, due to the lack of true concurrent processing, represents another limitation of the current system.

{The available protocol set for data access, namely redundant and failure protocols, are introduced to achieve the two extreme operating points of the trade-off between the efficiency and latency performance of tape libraries. While the redundant protocol has the potential to achieve better data access latencies in absence of heavy load, the failure protocol on the other hand, targets specifically heavy traffic and attempts at reducing the number of requests imposed on the double queues. Essentially, this is highlighting that  better optimized protocols can be implemented that could adapt to the workload patterns and achieve a dynamic compromise  between efficiency and latency dimensions. However, development of such protocols are library topology and architecture dependent, and building an adaptive scheme might require implementation of complex decision making processes, which are currently not supported  by our simulator.} 

{Finally, the complete interplay between the erasure coding and deduplication is not implemented in our simulator as this subject is still under active investigation by many researchers and hard to tackle once for all scenarios \cite{arslan2017joint}. For instance, it is imperative to address the fusion details concerning the generation of redundancy and data compression, particularly data with intrinsic redundancy, prior to embarking on any implementation efforts. Despite its complexity, understanding and resolving these intricacies are important to ensure the accuracy and efficiency of the simulated system.}

\vspace{3mm}

\section{Conclusion}

In this study, we have developed and presented a discrete-event tape simulation platform that incorporates realistic system-level assumptions, facilitating the accurate estimation of key performance indicators (KPIs) for both Enterprise and RAIL-type ML system configurations. The utilization of two types of redundancy, namely replication and erasure coding, ensures reliable data storage in a networked tape cloud setting. Through a constrained exploration of the parameter space, we compared scale-up (single large Enterprise-grade) and scale-out (RAIL-type, involving multiple smaller and less expensive) libraries. Our findings indicate the superiority of scale-out libraries due to their ability to shard traffic, efficiently utilizing library resources through inherent parallelism. Given the carefully engineered mechanics and complex operational aspects of tape systems, traditional queue theory fundamentals are insufficient for their analysis. Consequently, {despite its limitations with regards to data requests, internal mechanism details and the way data parallism is implemented,} the presented simulation platform {is proposed to serve} as a design tool for reliability engineering. {Essentially, the proposed tool} provides {quick} guidance {and simplified implementation best practices} to system administrators seeking to leverage dynamic libraries for long-term archival and data retention purposes.

%A discrete--event tape simulation platform is developed and presented with realistic system-level assumptions in which many key performance indicators (KPIs) can be accurately estimated for Enterprise and RAIL-type ML system configurations. Two types of redundancy, namely replication and erasure coding, are utilized to provide reliable data storage in a networked tape cloud setting. With very limited sampling from the parameter space, we have been able to compare scale-up (single large Enterprise-grade) and scale-out (RAIL-type multiple smaller and less expensive) libraries and conclude the scale-out libraries are more advantageous to use due to sharding the traffic, efficiently using the library resources due its inherent parallelism. Due to complex mechanics and complicated system operation, tape systems cannot easily be analyzed with known queue theory fundamentals. Thus, the presented simulation platform will serve as a guiding design tool for reliability engineering as well as system admins whom wants to utilize tape libraries for long term archive and data retention option. 

\section*{Appendix: Artifact Description/Evaluation}

\section*{\large Summary of the experiments reported}

In our experiments, we extensively explored different parameter settings (1) to investigate normal library operations as well as libraries under heavy load and (2) limitations of library systems, where they can go unstable upon the unsupervised selections of parameters. A simple way to simulate unstable system operation is to manually change the rate of user requests ($\lambda$), names as  
\texttt{p\_lam\_per\_day}. Our simulation platform produces multiple plots including latency of libraries (in mins) and user data requests and robot exchanges both as a function of simulation time (in terms of days). We also plot read errors as a function of hourly time steps. 

The baseline inputs to the simulator include library fill rate (percent), total number of cartridges per library, total number of robots per library, total number of drives per library, robot to drive time (secs), total annual touch rate, average object size (MB) and distribution, simulation time (months), simulation step size (secs), cartridge capacity (TB), ECC code used (k/n), number of libraries, Drive read failure probability, collocation parameter (MB), protocol used and whether robots (if more than one exists) are used in balanced fashion or not. The outputs of the simulator include total capacity of the system in simulation (PB), total number of objects touched, actual robot exchange rate, maximum, minimum and average of data access (both for the first byte or the last byte) latency (mins), user object request rate, distribution of robot movements, distributions and average of queue lengths for both DR and R queues. 

Our software consists of two parts. The first part generates a csv file such as \texttt{simQ.csv}. The csv file is processed separately after the main body of the simulation is over by the second part. The content of csv file includes columns \textsf{QID} $\in \{ DR, R\}$, \textsf{Q\_in}, \textsf{D\_in}, \textsf{Q\_out}, \textsf{D\_out}, \textsf{Q\_len} and \textsf{Data\_out} whereas each row corresponds to a event stored in one of the queues. The time step index is recorded in each one of the columns to indicate when a particular event starts and ends. Here \textsf{Q\_in}, \textsf{D\_in}, \textsf{Q\_out} values correspond to checkpoints Q-in, Data-in, Q-out   as mentioned in subsection 2.2.4. In addition, \textsf{Q\_len} keeps the queue length and \textsf{D\_out} corresponds to Data-access checkpoint. 

While executing the simulator with a redundancy mechanism such as replication or erasure coding, we also need to keep track of message IDs (\textsf{MID}) to be able to differentiate between different copies of different fragments of the same data block. For instance 2nd copy of the the block 312 is recorded as 312.2. 

In the case of RAIL-type ML simulations, we run multiple and concurrent instances of the software and record csv files with the following naming conventions \texttt{simQ0.csv}, \texttt{simQ1.csv}, \texttt{simQ2.csv}, $\dots$. The correlations are driven between multiple files by processing them based on the message IDs and recorded time stamps. 

\section*{\large Implementation}

Talics$^3$ simulator  is implemented in Python 2.7.15. We run our experiments both on Windows and Linux operating systems (OS). For windows OS, we have used PyCharm community edition 2016.3.2. Java Runtime Environment is selected to be 1.8.0\_112-release-408-b6 and the OpenJDK java virtual machine by JetBrains is supported. We have also tested the same software on Ubuntu LTS 20.4 OS. We have run both OSes on Intel i5-5300 CPU with four cores operating at 2.3GHZ each. In addition, for a proper compilation and runtime, the current version may require a Linux runtime environment and a C compiler (e.g. gcc version $\geq 4.8.4$). 

Our software utilizes many Python packages as dependencies: beautifulsoup4 4.9.0, gspread 3.4.2, httplib2 0.17.2, kiwisolver 1.0.1, lxml 4.5.0, matplotlib 2.2.2, numpy 1.14.5, pandas 0.24.2, pip 9.0.3, python-dateutil 2.7.3, requests 2.23.0, scipy 1.2.2, setuptools 44.1.0, urllib3 1.25.8.




\bibliographystyle{elsarticle-num}
\bibliography{cas-refs}

\begin{thebibliography}{10}
\expandafter\ifx\csname url\endcsname\relax
  \def\url#1{\texttt{#1}}\fi
\expandafter\ifx\csname urlprefix\endcsname\relax\def\urlprefix{URL }\fi
\expandafter\ifx\csname href\endcsname\relax
  \def\href#1#2{#2} \def\path#1{#1}\fi

\bibitem{IDC2025}
D.~Reinsel, J.~Gantz, J.~Rydning, White paper: The digitization of the world-from edge to core, Tech. rep., Technical Report US44413318, International Data Corporation, Framingham~… (2018).

\bibitem{gantz2011extracting}
J.~Gantz, D.~Reinsel, et~al., Extracting value from chaos, IDC iview 1142~(2011) (2011) 1--12.

\bibitem{Moore}
R.~L. Moore, J.~D'Aoust, R.~H. McDonald, D.~Minor, Disk and tape storage cost models, in: Archiving Conference, Vol. 2007, Society for Imaging Science and Technology, 2007, pp. 29--32.

\bibitem{mardis2011decade}
E.~R. Mardis, A decade’s perspective on dna sequencing technology, Nature 470~(7333) (2011) 198--203.

\bibitem{Appuswamy2019}
R.~Appuswamy, K.~Le~Brigand, P.~Barbry, M.~Antonini, O.~Madderson, P.~Freemont, J.~McDonald, T.~Heinis, Oligoarchive: Using dna in the dbms storage hierarchy., in: CIDR, 2019.

\bibitem{Bornholt2016}
J.~Bornholt, R.~Lopez, D.~M. Carmean, L.~Ceze, G.~Seelig, K.~Strauss, A dna-based archival storage system, in: Proceedings of the Twenty-First International Conference on Architectural Support for Programming Languages and Operating Systems, 2016, pp. 637--649.

\bibitem{gupta2012enabling}
R.~Gupta, H.~Gupta, U.~Nambiar, M.~Mohania, Enabling active data archival over cloud, in: 2012 IEEE Ninth International Conference on Services Computing, IEEE, 2012, pp. 98--105.

\bibitem{memishi}
B.~Memishi, R.~Appuswamy, M.~Paradies, Cold storage data archives: More than just a bunch of tapes, in: Proceedings of the 15th International Workshop on Data Management on New Hardware, 2019, pp. 1--7.

\bibitem{cancio2015experiences}
G.~Cancio, V.~Bahyl, D.~F. Kruse, J.~Leduc, E.~Cano, S.~Murray, Experiences and challenges running cern's high capacity tape archive, in: Journal of Physics: Conference Series, Vol. 664, IOP Publishing, 2015, p. 042006.

\bibitem{pease}
D.~Pease, A.~Amir, L.~V. Real, B.~Biskeborn, M.~Richmond, A.~Abe, The linear tape file system, in: 2010 IEEE 26th Symposium on Mass Storage Systems and Technologies (MSST), IEEE, 2010, pp. 1--8.

\bibitem{7877100}
M.~Mäsker, L.~Nagel, T.~Süß, A.~Brinkmann, L.~Sorth, Simulation and performance analysis of the ecmwf tape library system, in: SC '16: Proceedings of the International Conference for High Performance Computing, Networking, Storage and Analysis, 2016, pp. 252--263.
\newblock \href {https://doi.org/10.1109/SC.2016.21} {\path{doi:10.1109/SC.2016.21}}.

\bibitem{szor2005art}
P.~Szor, The art of computer virus research and defense: Art comp virus res defense \_p1, Pearson Education, 2005.

\bibitem{IPFS}
J.~Benet, Ipfs-content addressed, versioned, p2p file system, arXiv preprint arXiv:1407.3561 (2014).

\bibitem{arslan2022compress}
S.~S. Arslan, T.~Goker, Compress-store on blockchain: a decentralized data processing and immutable storage for multimedia streaming, Cluster Computing 25~(3) (2022) 1957--1968.

\bibitem{kiemle2016}
S.~Kiemle, K.~Molch, S.~Schropp, N.~Weiland, E.~Mikusch, Big data management in earth observation: The german satellite data archive at the german aerospace center, IEEE Geoscience and remote sensing magazine 4~(3) (2016) 51--58.

\bibitem{arslan2014mds}
S.~S. Arslan, J.~Lee, J.~Hodges, J.~Peng, H.~Le, T.~Goker, Mds product code performance estimations under header crc check failures and missing syncs, IEEE Transactions on Device and Materials Reliability 14~(3) (2014) 921--930.

\bibitem{meisling1958discrete}
T.~Meisling, Discrete-time queuing theory, Operations Research 6~(1) (1958) 96--105.

\bibitem{9614293}
I.~Iliadis, L.~Jordan, M.~Lantz, S.~Sarafijanovic, Performance evaluation of automated tape library systems, in: 2021 29th International Symposium on Modeling, Analysis, and Simulation of Computer and Telecommunication Systems (MASCOTS), 2021, pp. 1--8.
\newblock \href {https://doi.org/10.1109/MASCOTS53633.2021.9614293} {\path{doi:10.1109/MASCOTS53633.2021.9614293}}.

\bibitem{Ford1996}
D.~A. Ford, R.~J. Morris, A.~E. Bell, Redundant arrays of independent libraries (rail): a tertiary storage system, in: COMPCON'96. Technologies for the Information Superhighway Digest of Papers, IEEE, 1996, pp. 280--285.

\bibitem{Zeng1}
L.~Zeng, D.~Feng, F.~Wang, Z.~Cheng, Q.~Zou, Creation of a java-based interactive modeling environment with tape library model example, in: 2006 International Workshop on Networking, Architecture, and Storages (IWNAS'06), IEEE, 2006, pp. 8--pp.

\bibitem{7774566}
I.~{Iliadis}, Y.~{Kim}, S.~{Sarafijanovic}, V.~{Venkatesan}, Performance evaluation of a tape library system, in: 2016 IEEE 24th International Symposium on Modeling, Analysis and Simulation of Computer and Telecommunication Systems (MASCOTS), 2016, pp. 59--68.
\newblock \href {https://doi.org/10.1109/MASCOTS.2016.37} {\path{doi:10.1109/MASCOTS.2016.37}}.

\bibitem{arslan2020data}
S.~S. Arslan, J.~Peng, T.~Goker, A data-assisted reliability model for carrier-assisted cold data storage systems, Reliability Engineering \& System Safety 196 (2020) 106708.

\bibitem{masker2016simulation}
M.~M{\"a}sker, L.~Nagel, T.~S{\"u}{\ss}, A.~Brinkmann, L.~Sorth, Simulation and performance analysis of the ecmwf tape library system, in: SC'16: Proceedings of the International Conference for High Performance Computing, Networking, Storage and Analysis, IEEE, 2016, pp. 252--263.

\bibitem{tgau}
J.~L{\"u}ttgau, Modeling and simulation of tape libraries for hierarchical storage mangement systems, Ph.D. thesis, Universit{\"a}t Hamburg, Fachbereich Informatik (2016).

\bibitem{Lavengerg}
S.~S. Lavenberg, D.~R. Slutz, Regenerative simulation of a queuing model of an automated tape library, IBM Journal of Research and Development 19~(5) (1975) 463--475.

\bibitem{arslan2023durability}
S.~S. Arslan, Durability and availability of erasure-coded storage systems with concurrent maintenance, arXiv preprint arXiv:2301.09057 (2023).

\bibitem{fishman1978}
B.~Pokhodzei, Principles of discrete event simulation: Gs fishman. xviii+ 514 p. wiley-interscience, new york, 1978 (1981).

\bibitem{luttgau2016modeling}
J.~L{\"u}ttgau, Modeling and simulation of tape libraries for hierarchical storage mangement systems, Ph.D. thesis, Universit{\"a}t Hamburg, Fachbereich Informatik (2016).

\bibitem{Satyanarayanan}
M.~Satyanarayanan, A study of file sizes and functional lifetimes, ACM SIGOPS Operating Systems Review 15~(5) (1981) 96--108.

\bibitem{downey}
A.~B. Downey, The structural cause of file size distributions, in: MASCOTS 2001, Proceedings Ninth International Symposium on Modeling, Analysis and Simulation of Computer and Telecommunication Systems, IEEE, 2001, pp. 361--370.

\bibitem{arslan2020distribution}
S.~S. Arslan, E.~Zeydan, On the distribution modeling of heavy-tailed disk failure lifetime in big data centers, IEEE Transactions on Reliability 70~(2) (2020) 507--524.

\bibitem{gong}
W.~Gong, Y.~Liu, V.~Misra, D.~Towsley, On the tails of web file size distributions, in: Proceedings of the annual allerton conference on communication control and computing, Vol.~39, The University; 1998, 2001, pp. 192--201.

\bibitem{plank2008jerasure}
J.~S. Plank, S.~Simmerman, C.~D. Schuman, Jerasure: A library in c/c++ facilitating erasure coding for storage applications-version 1.2, University of Tennessee, Tech. Rep. CS-08-627 23 (2008).

\bibitem{arslan2021founsure}
{\c{S}}.~{\c{S}}. Arslan, Founsure 1.0: An erasure code library with efficient repair and update features, SoftwareX 13 (2021) 100662.

\bibitem{wicker1999reed}
S.~B. Wicker, V.~K. Bhargava, Reed-Solomon codes and their applications, John Wiley \& Sons, 1999.

\bibitem{Ross2014}
S.~M. Ross, Introduction to Probability Models, ISE, Academic press, 2006.

\bibitem{arslan2017joint}
S.~S. Arslan, T.~Goker, R.~Wideman, A joint dedupe-fountain coded archival storage, in: 2017 IEEE International Conference on Communications (ICC), IEEE, 2017, pp. 1--7.

\end{thebibliography}


\end{document}